\documentclass[conference]{IEEEtran}
\IEEEoverridecommandlockouts

\usepackage{cite}
\usepackage{hyperref}
\usepackage{amsmath,amssymb,amsfonts}
\usepackage{algorithmic}
\usepackage{graphicx}
\usepackage{textcomp}
\usepackage{xcolor}
\usepackage{tabularx}
\usepackage{booktabs}
\usepackage{comment} %
\usepackage{tikz} %
\usetikzlibrary{shapes.multipart, arrows.meta, positioning}
\ifCLASSOPTIONcompsoc
\usepackage[caption=false,font=normalsize,labelfon
t=sf,textfont=sf]{subfig}
\else
\usepackage[caption=false,font=footnotesize]{subfi
g}
\fi

\def\BibTeX{{\rm B\kern-.05em{\sc i\kern-.025em b}\kern-.08em
    T\kern-.1667em\lower.7ex\hbox{E}\kern-.125emX}}
\begin{document}

\title{Cross-Organizational Analysis of Parliamentary Processes: A Case Study
}

\author{
\IEEEauthorblockN{
Paul-Julius Hillmann\IEEEauthorrefmark{1},
Stephan A. Fahrenkrog-Petersen\IEEEauthorrefmark{3},
Jan Mendling\IEEEauthorrefmark{1}\IEEEauthorrefmark{2}
}
\\
\IEEEauthorblockA{\IEEEauthorrefmark{1}Humboldt-Universität zu Berlin, Berlin, Germany \\
\{paul.hillmann, jan.mendling\}@hu-berlin.de}
\IEEEauthorblockA{\IEEEauthorrefmark{2}Weizenbaum Institute, Berlin, Germany}
\IEEEauthorblockA{\IEEEauthorrefmark{3}University of Liechtenstein, Vaduz, Liechtenstein
\\
stephan.fahrenkrog@uni.li}
}

\maketitle

\begin{abstract}
Process Mining has been widely adopted by businesses and has been shown to help organizations analyze and optimize their processes. 
However, so far, little attention has gone into the cross-organizational comparison of processes, since many companies are hesitant to share their data. 
In this paper, we explore the processes of German state parliaments that are often legally required to share their data and run the same type of processes for different geographical regions.
This paper is the first attempt to apply process mining to parliamentary processes and, therefore, contributes toward a novel interdisciplinary research area that combines political science and process mining.
In our case study, we analyze legislative processes of three German state parliaments and generate insights into their differences and best practices. We provide a discussion of the relevance of our results that are based on knowledge exchange with a political scientist and a domain expert from the German federal parliament. 
\end{abstract}


\section{Introduction}\label{sec:intro}
Cross-organizational process mining~\cite{rott2024laying} has been presented as a group of techniques that help to compare similar processes from different organizations with the aim of facilitating benchmarking and mutual learning.
Many common processes, such as purchase-to-pay or order-to-cash, are operated by multiple organizations.
Understanding how an organization can learn best practices from another organization is a challenging and important question~\cite{abb2025identifying}. This idea of identifying best practices and how well an organization is run is commonly referred to as benchmarking.

Benchmarking is an established field of business research. Many of the approaches to benchmarking have long focused on black-box input-output models, with few examples taking processes directly into account~\cite{teuteberg2013semantic}.
Process mining has been positioned as a family of techniques to facilitate a detailed comparison as an aid for benchmarking.
So far, research on process mining has approached this application area as a technical challenge and provided technical contributions. At this stage, we know little about the effectiveness of process mining for such cross-organizational scenarios~\cite{rott2024laying}.

In this paper, we formulate the research objective to obtain empirical insights into the effectiveness of process mining for cross-organizational comparison. To that end, we make use of the domain of parliamentary processes and a case study research design. More specifically, we analyze the legislation processes of three German state parliaments and examine to which extent process mining techniques can provide more detailed insights into the relative productivity of state parliaments than the benchmark from political science research to measure relative legislative capacity \cite{Appeldorn.2022}.
In this way, we provide an empirical contribution to understanding the challenges of applying process mining in a cross-organizational setting and the insights that it provides in contrast to traditional measures applied in the domain. 
We found that process mining techniques can effectively be used in practice to reveal and explain differences in operational details such as the durations of processes.
Such empirical knowledge can inform future technical work on new algorithmic approaches and opens the door for collaborations between the process mining community and political scientists.

The remainder of the paper is structured as follows: \autoref{sec:back} introduces the relevant background. In \autoref{sec:method}, we explain our research method. Next, we present the results of the cross-organizational process analysis in \autoref{sec:analysis-results}. Finally, we conclude the paper in \autoref{sec:conclusion}.

\section{Background}\label{sec:back}

In this section, we are going to introduce the relevant background to our work. In \autoref{sec:case_setting}, we describe the background of our case. Next, in \autoref{sec:prior_research}, we describe relevant prior research in the field of cross-organizational process mining. 

\subsection{Case Setting}
\label{sec:case_setting}
As a federal republic, Germany consists of 16 states (also known as \emph{Bundesländer}) that have their own parliament, responsible for the same tasks in distinct geographic locations. 
German state parliaments offer an interesting case to perform a cross-organizational analysis~\cite{rott2024laying} for several reasons: (i) all state parliaments perform the same type of processes; (ii) some parliaments share their data fully as open data; (iii) the quality of the data is a high priority for the organizations. 
Further, so far, no study in the process mining literature has worked with parliamentary processes and this topic opens the door to collaboration between the process mining community and political scientists.
Other communities, such as the NLP community~\cite{ReinigRP24} already utilize the public data of German parliaments for their research. 

The German state parliaments are required to document their work through different documents such as session protocols or law drafts. These respective documents are organized by the documentation service of each parliament.
Often, the different documents belonging to the same procedure, i.e., passing of a law, are linked together.  
For this study, we manually investigated the website of each documentation service to assess whether they publish machine-readable versions, such as XML files, of their documentation. A total of three states (Berlin, Brandenburg, and Saxony) have these datasets readily available for download. However, the data by the parliament of Saxony was not usable for our purposes since it included only very rudimentary information, showing no attributes corresponding to the date of the documents or any hint towards connections among them.
We further contacted all remaining 13 states via phone calls and e-mails to inquire about the availability of such data. The state of Baden-Württemberg provided us with data. 
In sum, we obtained data from the state parliaments of Berlin, Brandenburg, and Baden-Württemberg: for each, one XML file per election period. The data for all three roughly goes back to years around the German unification in 1990 and runs until the day of our download. For Berlin, the data goes back to their 11th election period starting in 1989; for Brandenburg, it goes back to their first election period starting in 1990; and for Baden-Württemberg, the obtained data starts with their 9th election period in 1984.
\begin{figure}[t]
    \centering
    \includegraphics[width=1\linewidth]{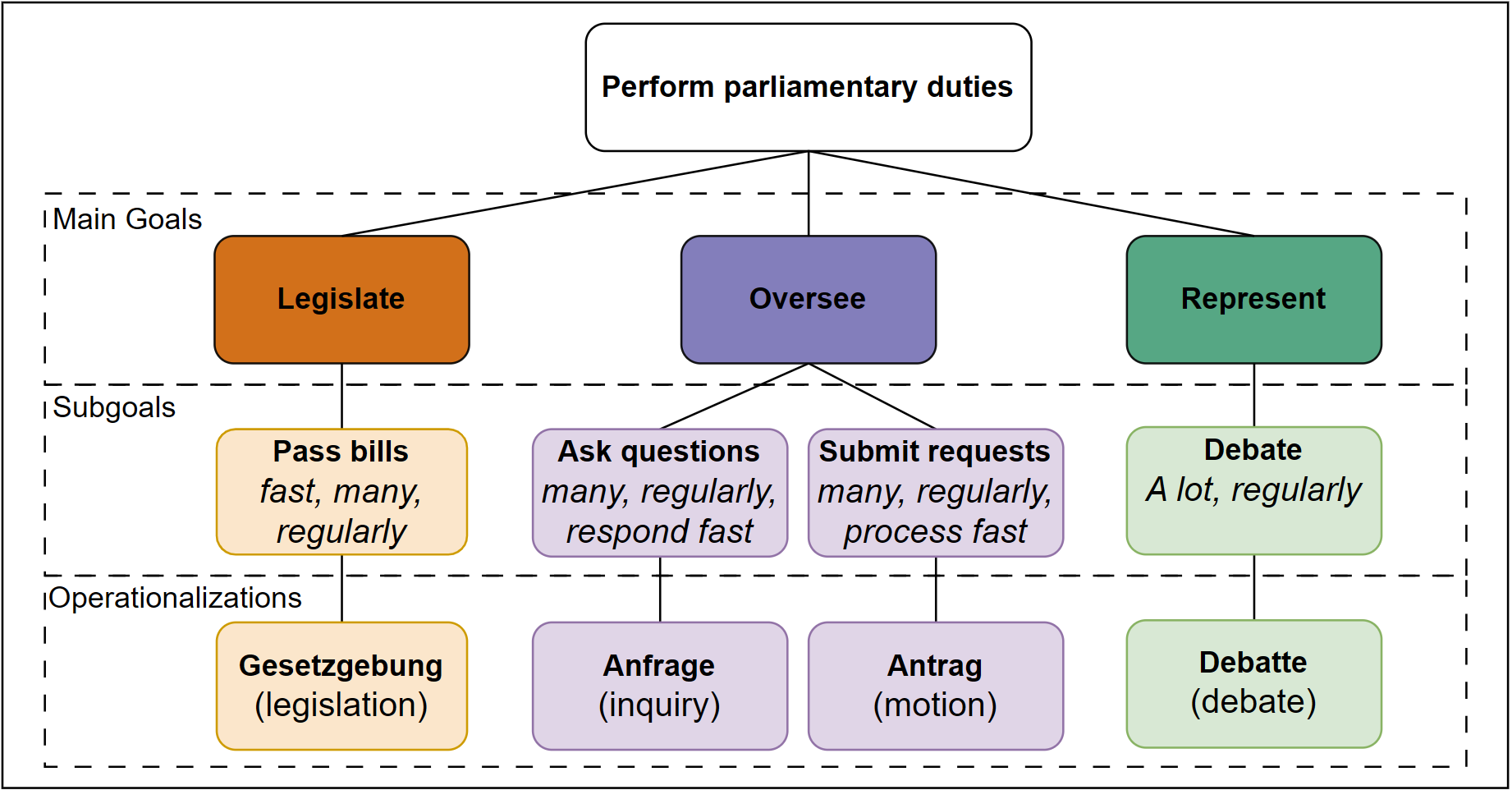}
    \hfill        
    \caption{The goal-based process architecture, showing the most relevant identified processes as operationalizations of parliamentary goals.}
    \label{fig:processArchitectureCompact}
\end{figure} 

To better understand the business processes of state parliaments, we constructed a goal-based process architecture, as shown in ~\autoref{fig:processArchitectureCompact}. 
Though different lists of functions of parliaments have been proposed \cite{Aldons.2001}, generally the main goals of parliaments can be summarized as: (i) to \emph{Legislate}; (ii) \emph{Oversee} the government, and (iii) \emph{Represent} the electorate \cite{Kinyondo.2022}.
The goals can be broken down into more concrete subgoals: To legislate, bills should be passed. To oversee, the government should be asked questions and requests should be submitted. Furthermore, to represent, there should be debates.

To quantify these goals, we postulated the following desired process outcomes related to these goals:
\begin{itemize}
    \item Pass bills: fast, many, regularly 
    \item Ask questions and submit requests: many, regularly - and answer or process the questions and requests quickly
    \item Debate: a lot, regularly 
\end{itemize}
These outcomes are simplified, but they allow us to assume desired properties of the underlying processes.
These properties are close enough to real-world goals of parliaments for the purposes of our case study. 

The comparison of the capacity of state parliaments to meet their goals has a long history in political science. The Squire Index \cite{Squire.1992} can be considered the most relevant benchmark for measuring and comparing the legislative capacity of US state parliaments \cite{brown2025comparing}. Recently, it was also used in research on German parliaments, demonstrating its broader applicability and usefulness \cite{Appeldorn.2022}. The Squire Index defines the legislative capacity of state parliaments by considering the legislators' salary, staff support, and the parliaments' session lengths in relation to the same measures for the corresponding national parliament. The assumption is that more work can be done in the parliament if these properties are higher: Legislators with higher pay can focus more of their time on parliamentary work, and higher staff support and more or longer parliamentary sessions allow for more output. A large variety of studies have used the Squire Index as an explanatory variable, showing many impacts of the expressed legislative capacity, including impacts on the ways in which parliaments operate and on the types of proposed and adopted laws \cite{Squire.2024}.

\subsection{Prior Research}
\label{sec:prior_research}
The Squire Index is an important example of a benchmark.
Benchmarking is an important tool for businesses to identify best practices for excellence in products, services, and processes through comparisons to other organizations \cite{bhutta1999benchmarking, talluri2000benchmarking}. Therefore, it can serve as a trigger for improvement and innovation \cite{anand2008benchmarking}. Generally, benchmarking can be performed in a quantitative way focusing on performance indicators or in a qualitative way, examining performance gaps more in-depth \cite{teuteberg2013semantic}. Qualitative benchmarking allows not only for comparisons, but aims at uncovering the causes of performance differences - traditionally through techniques like interviewing, document analysis, or observations \cite{teuteberg2013semantic}.

The process mining community has contributed to facilitating benchmarking through so-called cross-organizational process mining exploiting commonalities - i.e., the application of process mining techniques in a setting where multiple organizations perform variants of the same process \cite{vanderAalst_intraandinterOrga.2011}. However, even though this type of process mining was recognized as one of the key research challenges in the process mining manifesto \cite{vanderAalst_manifesto.2012} more than ten years ago, and despite subsequent technological advances, ongoing calls for further research, and organizations showing high interest in it, the practical adoption of cross-organizational process mining remains very limited \cite{rott2024laying}. In a recent literature review, Rott et al. \cite{rott2024laying} found that research in this area has predominantly concentrated on technical aspects with a fragmented body of work, resulting in only limited empirical insights on how effective these techniques actually are in practice. Notably, they found no published articles on cross-organizational process mining exploiting commonalities since 2019. So far, most of the existing work mainly used very few datasets from Australian hospitals \cite{Suriadi.2014, partington.2015, pini.2015} and Dutch municipalities \cite{Buijs.2012, buijs.2013, Buijs.2014, bernardi.2014, Yilmaz.2015}.

\section{Research Method}
\label{sec:method}
Methodologies to generate new knowledge can be grouped into knowledge \textit{of} and knowledge \textit{about} tasks and designs \cite{mendling2023methodology}. For instance, knowledge \textit{of} algorithm design refers to how an algorithm works, while knowledge \textit{about} algorithm design includes insights about its characteristics like its performance - e.g., to which extent an algorithm satisfies the task requirements. 
In this work, the objective is not to create new knowledge \textit{of} algorithm design, but to inductively generate new knowledge \textit{about} process mining techniques with respect to their effectiveness in real-world cross-organizational scenarios. We pursue this objective through a positivist confirmatory case study research design using archival data \cite{runeson2009casestudies}. Thereby, we examine the often implicit hypothesis that process mining techniques are indeed effective in real-world benchmarking settings. The data we collected from the three German state parliaments, and our case study design, are explained in the following.

\subsection{Source Data}

All the collected data files share a common structure. This structure is explained in a document available on the website of the state parliament of Berlin~\cite{AbgeordnetenhausBerlin.2024}. However, this file only outlines the basic structure of the files and the attributes that may exist, without actually explaining any semantics - no additional documentation was found. For additional explanations about the data, we consulted employees of the parliamentary documentation service of Berlin. 
The general structure of the raw data adheres to the following principles: The outermost element is always called \textit{Export}. It can contain any number of elements of the name \textit{Vorgang} (\textit{process}). These process elements can have different attributes: Besides internal information like an identifier, the most interesting attributes are a type attribute (short: \textit{VTyp}, long: \textit{VTypL}), and a so-called \textit{Systematik} (short: \textit{VSys}, long: \textit{VSysL}), which is a typical attribute used in libraries for organizational and classification purposes. Then, each process element can contain any number of elements called \textit{Nebeneintrag} (\textit{side entry}) - these typically contain a descriptor for the process. Most importantly, the process elements contain \textit{Dokument} (\textit{document}) elements. These are the elements representing the documents that can be searched for and downloaded on the documentation services' websites. Document elements can have several attributes: In addition to some internal identifiers, each document element has a title, a date, a type, and a URL pointing to a PDF file. Furthermore, document elements can have attributes like descriptors, authors, speakers, or an abstract. 

Even though all the data files share this common structure, the possible values for all the attributes vary. This is obvious when talking for instance about the authors of a document. However, also attributes that could be the same across the parliaments are different - for example, the type properties of the documents show different possible values across the parliaments. The overall shared structure still allowed us to apply a common approach for transforming the raw data into event logs that could then be used as the input for process mining techniques.

\subsection{Event Log Generation}
From the obtained data, we generated one event log per state parliament in the standard XES format using the libraries Pandas \cite{pandas} and \textit{PM4PY} \cite{pm4py}. We used the following design decisions: 

\begin{enumerate}
    \item Case identifier: Each \textit{Vorgang} is treated as one trace of a process.
    \item Activity type: Each document is interpreted as an event. The document's type property \textit{DokTypL} encodes the corresponding activity.
    \item Timestamps: The document's date property \textit{DokDat} is used for the timestamp.
    \item Case attributes: The \textit{VSysL} property of the processes is integrated as a case attribute. 
    \item Event attributes:  Additionally, all attributes of the documents are integrated as event attributes in the event log.
    \item Sorting: To facilitate later comparisons and groupings by attributes, for all attributes that have lists as their values, the lists are sorted.
\end{enumerate}
The resulting event logs still had some data quality issues. We addressed timestamp issues in the following way:
\begin{itemize}
    \item Missing timestamps: All traces containing an event with a missing timestamp were removed.
    \item Temporal validity: All traces containing events with timestamps before 1984 or after 2024 were removed.
    \item Cycle time verification: All traces with an unrealistically high cycle time were also removed. Through a look at the data and considering the maximum length of election periods, which is five years, this maximum time was set to exactly that. This way also date-values that seem like proper values could be identified as invalid - for instance, a trace that started in 2001 and has one event that states the year 2011 can be identified as erroneous.
\end{itemize}

Second, concerning missing activity names - i.e., no entry for a document's property \textit{DokTypL} - our first idea was to remove all the traces containing at least one event that is missing a proper activity name. While for Berlin and Brandenburg, this seemed feasible as not many such traces exist in the data, for Baden-Württemberg the number of traces that would have to be removed seemed too great. Therefore, after more inspection, it became clear that in Baden-Württemberg it seems to be normal to have plenary protocols without a \textit{DokTypL} property. Next to these plenary protocols, also documents of law texts were found without a \textit{DokTypL} attribute. This information could be found in the document attribute \textit{DokArtL}.  For all three parliaments, most documents have the less informative value \textit{Drucksache} (\textit{printed matter}) for the \textit{DokArtL} attribute. However, in the cases where this value carries more information than just \textit{Drucksache}, and the \textit{DokTypL} attribute is missing, the \textit{DokArtL} attribute was used as the activity type instead.

In summary, the event log generation resulted in three pre-processed event logs - one per parliament. 
In \autoref{tab:event_logs_basicFacts}, we record the number of traces by each data transformation step and show basic properties of the three pre-processed event logs that were the basis for all following work. The event logs are unique in the process mining community, due to the extensive time period that they cover, because they cover the same type of processes in different organizations, and because of their origin in parliamentary processes. They can be found in a public repository\footnote{\url{https://github.com/pjhi/coPM-German-State-Parliament-Data}}.

\begin{table}[t]
\centering
\caption{Data overview for Berlin, Brandenburg, and Baden-Württemberg: trace filtering and properties of resulting event logs.}
\label{tab:event_logs_basicFacts}
\begin{tabular}{lccc}
    \toprule
    & Berlin & Brandenburg & Baden-Württemberg \\
    \midrule
    \multicolumn{4}{l}{\textbf{Trace Filtering:} No. of traces...} \\
    originally & $109\,370$ & $69\,740$ & $62\,393$ \\
    missing date & $4240$ & $60$ & $9$ \\
    invalid date & $10$ & $1$ & $23$ \\
    no activity name & $81$ & $8128$ & $27\,898$ \\
    \multicolumn{1}{l}{\parbox{2cm}{no activity name after correction}} 
            & $54$ & $78$ & $2912$ \\
    removed in total & $4279$ & $83$ & $2935$ \\
    after processing & $105\,091$ & $69\,657$ & $59\,458$ \\
    \midrule
    \multicolumn{4}{l}{\textbf{Event Log Properties}} \\
    Start & 18.11.1988 & 22.10.1990 & 01.06.1984 \\
    End & 23.08.2024 & 27.08.2024 & 29.08.2024 \\
    No. of events & $249\,229$ & $145\,469$ & $125\,981$ \\
    No. of activities & $71$ & $106$ & $153$ \\
    Median cycle time & $17$ days & $7$ days & $0$ days \\
    Mean cycle time & $50.1$ days & $29.5$ days & $87.5$ days \\
    \bottomrule
\end{tabular}
\end{table}

\subsection{Study}
Based on the generated event logs, the goal of this case study is to examine whether differences in the legislation processes of the different state parliaments can be appropriately analyzed.
We aim to identify differences in both the frequency of cases and the average cycle time.
We split the presentation of our case study into two parts: 
First, we will analyze how the Squire Index of the parliaments matches with our observed differences on the process level. 
Next, we analyze notable differences in the processes and potential root causes for them.

\subsubsection{Squire Index and Process Metrics}
For the purpose of this study, we focused on the lawmaking processes of the parliaments as an operationalization of the goal \textit{to legislate}. This process is also one of the most complex processes in the state parliaments.
We look at trace \emph{frequencies}, and \emph{cycle times} as proxies for our user-defined goals of
passing many bills quickly in order to legislate well.

We further observe the extent to which these proxy measurements correlate with the Squire Index. This allows us to check whether the actual process observations and the benchmark metric are consistent. 
We use the Pearson's correlation coefficient to assess the strength, direction, and significance of these correlations.
We consider correlations statistically significant if the $p$-value is below $0.05$.

\subsubsection{Analyzing Process Differences}
For this analysis, we first had to apply multiple pre-processing steps:
We filtered our event logs to only consider cases from 2006 to 2020. This is the latest 15-year window where data for the Squire Index is available.
We further pre-processed the activity labels to ensure that distinct activity labels exist. For example, German parliaments discuss bills in three so-called \emph{Readings}. These readings have a different procedure, but were partially encoded with the same activity label.
We fixed this by introducing distinct activity labels for every \emph{Reading} type.

Similar to Cremerius et al.~\cite{cremerius.2023}, we apply the \emph{Mann-Whitney U test} to assess whether the differences in cycle times between the organizations are statistically significant. The Mann-Whitney U test, as a non-parametric test, was particularly suitable because it does not rely on the assumption of an underlying normal distribution of the data.

Further, we use the dotted chart~\cite{vanderAalst.additionalPerspectives.dottedCharts.2016}, a visual analytics technique from process mining. This allows us to identify differences in the control flow of the different state parliaments. 
The dotted charts for this work were created using the implementation provided in the ProM framework (version 6.13) \cite{ProM6.Verbeek.2014} using the relative time option for the x-axis, sorted traces along the y-axis based on their cycle times, and colored the dots according to the activity types.

 Finally, we employ techniques from deviance mining and root cause analysis to pinpoint concrete reasons for differences in cycle time.
 To this end, we first have to determine what we consider a delayed case. We consider all traces delayed if and only if their cycle time exceeds the mean cycle time of the fastest state parliament (Baden-Württemberg) by more than 10\%.
 Applying this definition, in the data for Berlin and Brandenburg, around two thirds of all cases are delayed (Berlin: $483$, Brandenburg: $451$). In Baden-Württemberg, it is the opposite, only roughly around one third of the cases are delayed ($286$).
 We use the technique proposed by Völzer et al.~\cite{Volzer.2023} to generate potential root causes of the delay of traces. These root causes come in the form of inductive rules. 
 For model validation, a $0.33$ train-test split was used.
 For cases where the hypothesis induction returns rules that are not well interpretable or meaningful to the analyst, Völzer et al. \cite{Volzer.2023} propose several techniques - including \textit{rule simplification}, \textit{rule rewriting}, and \textit{attribute hiding}, which we also used in this work.
For instance, rules can manually be rewritten or simplified by removing conditions to make them more meaningful.
Additionally, through attribute hiding, the hypothesis inducer can be forced to find alternative rules without certain attributes.
Of course, after changing rules, their explanatory power has to be assessed again. Similarly to Völzer et al. \cite{Volzer.2023}, we assessed the quality of single rules by calculating their precision and recall. 

For this analysis, we enriched the event logs with additional features to allow for a better consideration of different factors.
First, we added contextual information to the event logs~\cite{Mendling.2020, Mikalef.2020}. This includes time-related features, such as if a year is an election year or the start month of a trace. Further, we added information about the general workload at the respective point in time to each trace.
We also added information about the first document of a trace (in our case, the bill under consideration), such as the size of the respective PDF file and its word count.
The political metrics related to the Squire Index~\cite{Appeldorn.2022} were also incorporated into the event logs. We further added a flag indicating whether the respective bill has been passed. 

We used two different settings for our approach to explain delays for both Berlin and Brandenburg: 
 \begin{enumerate}
    \item Hiding all time-related attributes to avoid circular explanations
    \item Allowing the delay attributes but hiding any attributes containing the word \textit{Gesetz} (law) - basically to avoid having rules that just use the delay from law initiative to passed law, and to force the inducer to focus on more detailed process steps
\end{enumerate}

Then, based on the automatically generated rules, we manually derived a set of comparable rules across the state parliaments by rewriting and simplifying to get an idea if they apply similarly for all three state parliaments or are actually a unique difference for one of them.

\section{Analysis Results}\label{sec:analysis-results}
In this section, we first introduce the results from our case study in \autoref{sec:results}.
Next, in \autoref{sec:discussion}, we discuss the implications of our work for the case study and future cross-organizational process mining research.

\subsection{Results}
\label{sec:results}
\subsubsection{High-level comparisons}
Based on our goal-based process architecture, we examined how the three state parliaments achieve their core democratic function to legislate through analyzing the yearly frequencies and mean cycle times of their lawmaking processes.
\begin{figure}[t]
\centering
    \includegraphics[width=\linewidth]{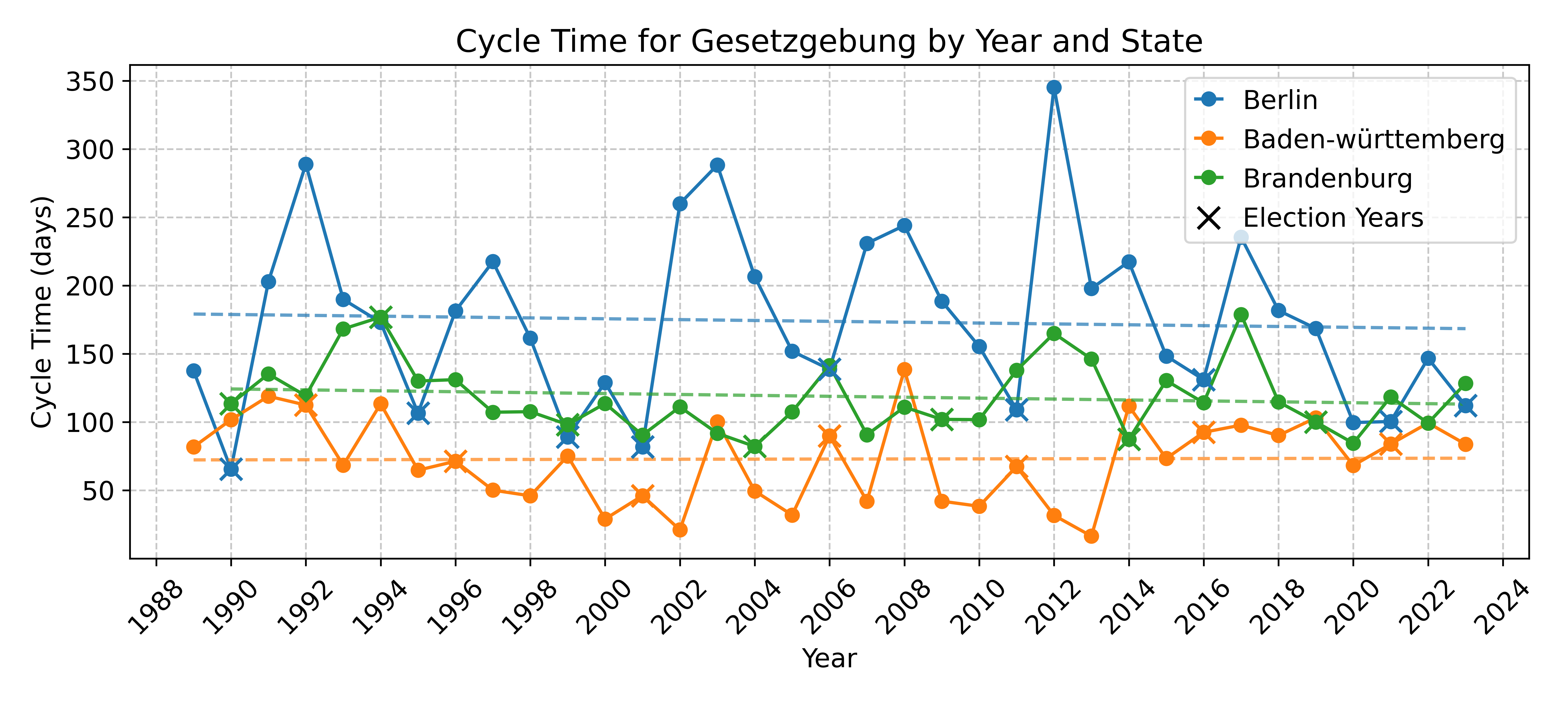}
    \caption{Mean cycle times of lawmaking processes per year and per parliament.}
    \label{fig:linePlots_lawmaking}
\end{figure}
\begin{table}[t]
    \centering
    \caption{Significant correlations between Squire Index and process metrics for legislation (Gesetzgebung) processes. Note: BE = Berlin, BB = Brandenburg, BW = Baden-Württemberg}
    \label{tab:squire_correlations_combined}
    \begin{tabular}{lllrrl}
        \toprule
        Metric & State & Time span & \textit{p}-value & Pearson $r$ & Expected? \\
        \midrule
                Frequencies & BE & 1998--2020 & 0.0325 & 0.4469 & Yes \\
                    & BB & 1991--2020 & 0.0095 & -0.4655 & No \\
                    & BW & 1990--2020 & 0.0055 & -0.4870 & No \\
        \bottomrule
    \end{tabular}
\end{table}

In terms of frequencies of the lawmaking processes per year, the numbers for Brandenburg and Berlin are more similar, while the numbers for Baden-Württemberg are different. The yearly process counts for the lawmaking process in Berlin and Brandenburg steadily move around $50$ over the whole time period. In contrast, the trend in Baden-Württemberg shows a decline starting around $150$ in early years and moving towards very similar values like in Berlin and Brandenburg from around 2011 until the end of the time period. Additionally, the numbers for Baden-Württemberg before 2011 show a lot of jumps between smaller and higher values.

\autoref{fig:linePlots_lawmaking} shows the yearly mean cycle times of the legislation processes in the three state parliaments. All three trend lines stay relatively stable over the whole time period, with the line for Berlin staying around $170$ days, the line for Brandenburg staying around $125$ days, and the line for Baden-Württemberg staying around $75$ days. However, the results for Berlin show the most variability, displaying a clear pattern regarding election years.

The assumption in this work is that with a higher Squire Index, more bills should be passed, and they should be passed more quickly as the legislative efficiency should be higher \cite{Squire.1998}. As can be seen in \autoref{tab:squire_correlations_combined}, contrary to expectations, Brandenburg and Baden-Württemberg show negative correlations between Squire Index values and process frequencies, indicating fewer legislation processes in years with higher legislative capacity. Only Berlin follows the expected pattern, with more frequent legislative activities as the Squire Index increases. Furthermore, no significant correlations were found for cycle times in any of the parliaments.

Interestingly, while we see differences in terms of the cycle time over time, we cannot explain them with the commonly used benchmark, the Squire Index. Therefore, we need to check in the next section if process mining gives us a better picture in this regard and if we gain actionable insights.

\subsubsection{In-depth lawmaking process analysis}
To gain a better understanding of the identified cycle time differences, we conducted a more detailed analysis. 
For this, we used pre-processed event logs that are described in \autoref{tab:basicFacts_initialEventLogs_legislation}.
Notably, Baden-Württemberg, the state parliament that on average showed the fastest legislation processes in the earlier analysis, shows the highest number of cases, the lowest mean number of events per case, the highest number of distinct activities, and together with Brandenburg a high number of variants.
Furthermore, the results for the Mann-Whitney U test for this time period confirm statistically significant differences between the cycle times of all states ($p < 0.001$).
\begin{table}[t]
    \centering
    \caption{Basic properties and information about cycle times for the three processed event logs for the legislation process for Berlin, Brandenburg, and Baden-Württemberg between 2006 and 2020}
    \label{tab:basicFacts_initialEventLogs_legislation}
    \begin{tabular}{lccc}
        \toprule
        & Berlin & Brandenburg & Baden-Württ. \\
        \midrule
        No. of cases & $731$ & $718$ & $1005$ \\
        No. of events & $5223$ & $6658$ & $6269$ \\
        \multicolumn{1}{l}{\parbox{2cm}{Mean no. of events per case}}  & $7.15$ & $9.27$ & $6.24$ \\ 
        No. of activities & $16$ & $24$ & $34$ \\
        No. of variants & $167$ & $378$ & $347$ \\
        \midrule
        Mean cycle time & $176.4$ days & $120.4$ days & $61.1$ days \\
        Median cycle time & $98.7$ days & $85$ days & $32$ days \\
        Std. deviation cycle time & $243.03$ days & $145.18$ days & $94.88$ days \\
        \bottomrule
    \end{tabular}
\end{table}

\begin{figure}[t]
\centering
    \subfloat[Berlin]{\includegraphics[width=\linewidth]{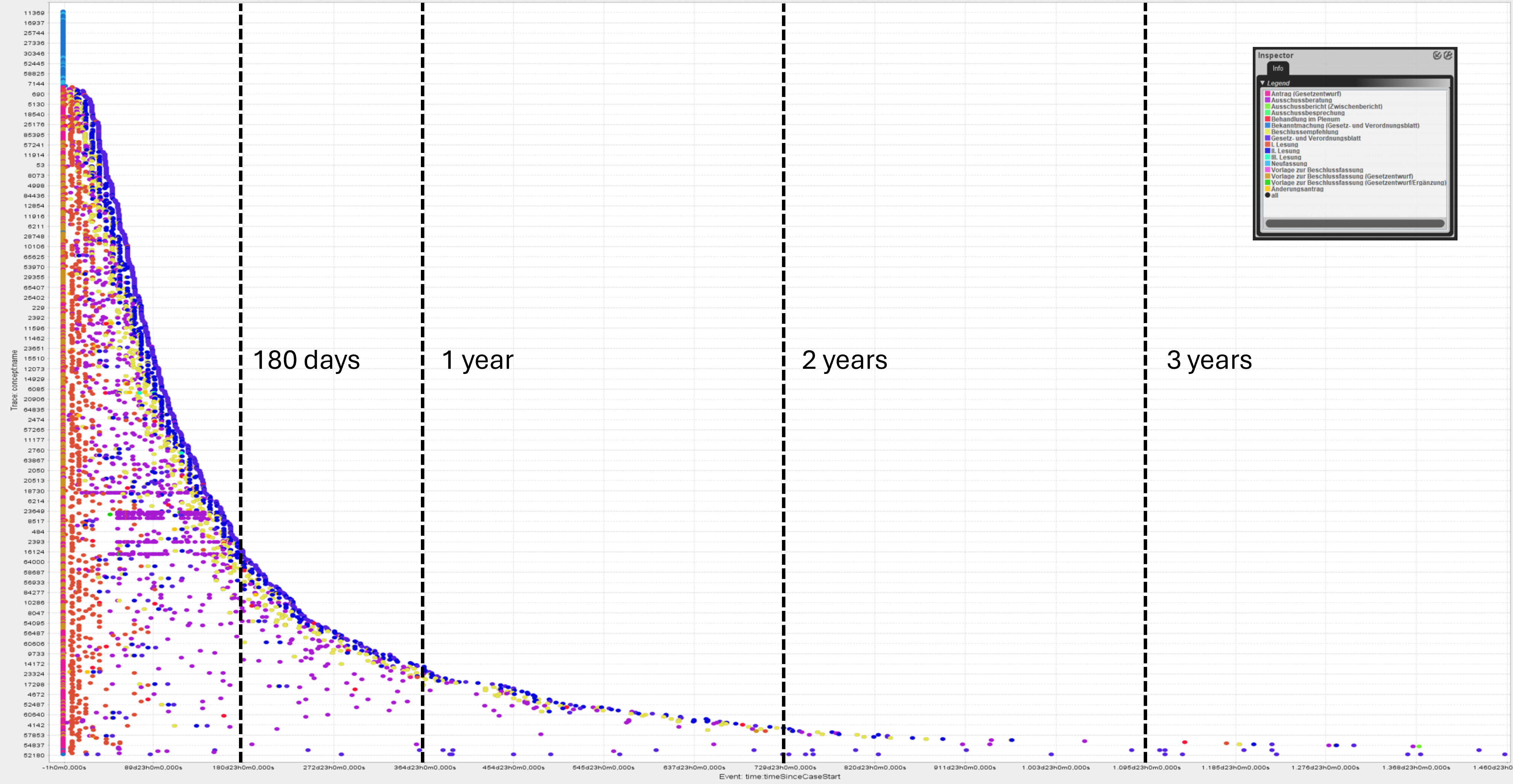}} \\
    \subfloat[Brandenburg]{\includegraphics[width=\linewidth]{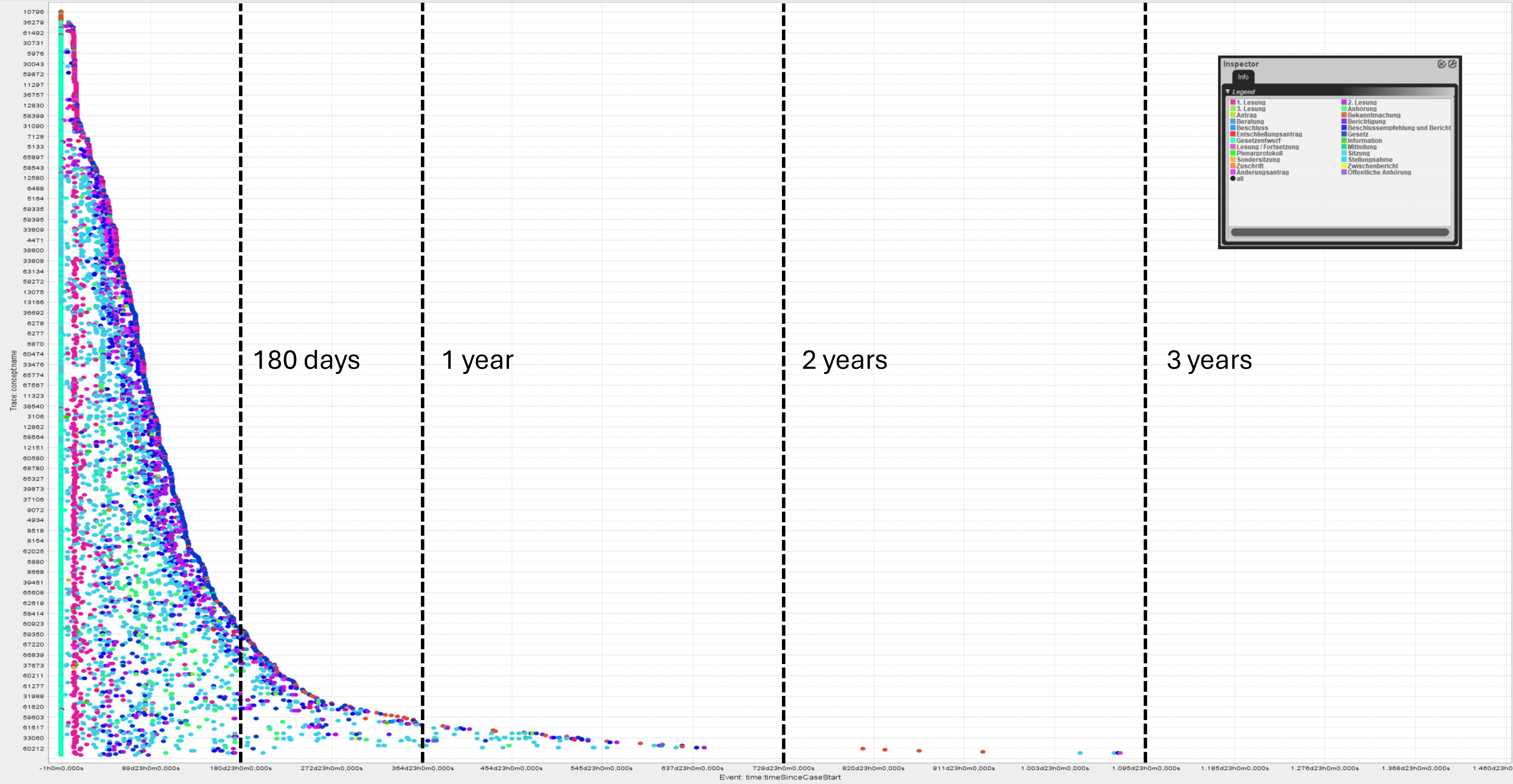}} \\
    
    \caption{Dotted charts showing 4 years of relative time along the x-axis, with all cases aligned to start at time 0 (a few dots for Berlin and Brandenburg fall outside this range). Each horizontal line along the y-axis represents a single case trace. The colors of dots indicate different activity types and do not match across parliaments.}
    \label{fig:allDottedCharts}
\end{figure}

Having established that the different state parliaments have significant differences in their cycle times, we turn to visualizing the control-flow of their legislation processes. In \autoref{fig:allDottedCharts}, we show two dotted charts, one for Berlin and one for Brandenburg; both states have longer cycle times than Baden-Württemberg.
The shown traces are sorted by their duration, therefore the different outer curves highlight the differences of the trace durations. 
It is immediately visible that Berlin has many long traces, with very long waiting periods between events. In contrast, long traces in Brandenburg have many events and significantly less waiting time between events.
Further, the color coding of the dotted charts gives additional clues for root causes of differences in cycle time.
For example, in the case of Berlin some traces stand out with many purple dots, indicating many committee consultations. Hinting that certain legislation might have taken a long time because of the need for many of such sessions.

Overall, the visual analysis enabled us to conclude that long waiting times between events can be a possible root cause for longer cycle times in Berlin. A similar conclusion cannot be made for Brandenburg. 
In contrast, Brandenburg shows significantly more procedural steps in the process.

Next, we turn to the induction of rules to explain the differences in cycle time. 
Based on the automatically generated rules and the knowledge obtained through the visual analysis, we manually derived five interpretable rules through simplification and rewriting, shown in \autoref{tab:rules_comparison}.
The following analyzes each rule set and its implications for understanding cycle time differences across the three state parliaments.

\begin{table}[t]
    \centering
    \caption{Comparison of rules derived from the different inductions for explaining delayed legislation processes across the three state parliaments Berlin, Brandenburg, and Baden-Württemberg.}
    \label{tab:rules_comparison}
    \footnotesize
    \begin{tabularx}{\columnwidth}{l X r r}
        \toprule
         & \textbf{Rule} & \textbf{Precision} & \textbf{Recall} \\
        \midrule
         & \textit{High Event Count} & & \\
        BE & event\_count $\geq$ 8.0 & 0.841 & 0.373 \\
        BB & event\_count $\geq$ 8.0 & 0.826 & 0.714 \\
        BW & event\_count $\geq$ 8.0 & 0.56 & 0.486 \\
        \midrule

         & \textit{Multiple Committee Session Activities} & & \\
        BE & Ausschussberatung.count $\geq$ 3.0 & 0.893 & 0.313 \\
        BB & Sitzung.count $\geq$ 3.0 & 0.804 & 0.672 \\
        BW & Beschlussempfehlung und Bericht.count $\geq$ 3.0 & 0.115 & 0.01 \\
        \midrule

         & \textit{Committee Session Delay After First Reading} & & \\
        BE & I. Lesung:Ausschussberatung.delay $\geq$ 24.0 & 0.961 & 0.762 \\
        BB & 1. Lesung:Sitzung.delay $\geq$ 24.0 & 0.961 & 0.324 \\
        BW & Erste Beratung:Beschlussempfehlung und Bericht.delay $\geq$ 24.0 & 0.882 & 0.42 \\
        \midrule

         & \textit{First to Second Reading Delay} & & \\
        BE & I. Lesung:II. Lesung.delay $\geq$ 30.0 & 0.934 & 0.903 \\
        BB & 1. Lesung:2. Lesung.delay $\geq$ 30.0 & 0.956 & 0.812 \\
        BW & Erste Beratung:Zweite Beratung.delay $\geq$ 30.0 & 0.816 & 0.528 \\
        \midrule

         & \textit{Passed Legislation} & & \\
        BE & is\_passed\_bill = True & 0.604 & 0.687 \\
        BB & is\_passed\_bill = True & 0.845 & 0.823 \\
        BW & is\_passed\_bill = True & 0.487 & 0.531 \\
        \bottomrule
    \end{tabularx}
\end{table}

\textbf{High event counts:}
This rule is motivated by a rule with a very high recall for Brandenburg that is in line with our learning from the dotted charts:
Traces with many events (more than 8) seem to need significantly longer than other traces. This rule has a very high precision in both Berlin and Brandenburg. Traces with many events could be related to bills that are either controversial or complex.

\textbf{Multiple committee sessions:}
This rule is a deviation of the previous rule. Rather than focusing on all types of activities, this rule focuses on committee sessions. If a trace has multiple committee sessions, it is also significantly more likely to be delayed in Berlin and Brandenburg. This finding was also supported by the dotted chart analysis for Brandenburg. 
This rule has unusually low support in Baden-Württemberg because the respective committee sessions are not included in the dataset. Instead, only decision recommendations and reports of the committees are included in the data, leading to a lower count for respective events. 
Therefore, the results for Baden-Württemberg have to be considered with a grain of salt with regard to this rule.

\textbf{Delay between first reading and committee session:}
The next rule states that a delay of 24 days or more between the first reading of a bill (its) and the first committee activity can result in a delay. This rule is supported in all three state parliaments. The rule fits well with the dotted chart analysis of Berlin.
Basically, this rule indicates that if between the introduction of a bill and the work on the bill a lot of time passes, it is likely that the trace will take a long time. The rule could cover traces that are of a low priority.

\textbf{Delay between first and second reading:}
Similar to the previous rule, this rule is concerned with the delay of the first and second reading of bills. If the delay is at least 30 days, we can explain many delays in all three state parliaments.
This confirms that the time between these key legislative milestones is a critical factor in determining overall process duration across all three parliaments.
 However, it is important to note that this rule tends toward a circular explanation: by definition, longer durations between major activities contribute directly to longer overall cycle times. Therefore, while statistically powerful, this rule provides less insight into root causes than our previous rules. Nevertheless, the lower recall for Baden-Württemberg suggests meaningful cross-organizational differences in how this particular delay impacts overall duration.

\textbf{Passed bills:}
The final rule examines if a bill was passed and how that impacts the cycle time. 
We created this rule manually based on an observation from the data of the Berlin parliament. By testing this rule on all parliaments, we see that it is moderately supported.
The intuition behind this rule is that if a bill is actually made into law, many resources and time are spent on improving the bill. If a law is likely to be rejected, e.g., because it is introduced by an opposition party, the bill will get less attention from the parliament.

\subsection{Discussion}
\label{sec:discussion}
Our case study demonstrates that cross-organizational process mining can be applied effectively in practice. Additionally, it shows that process mining techniques can be used to analyze and benchmark processes of parliaments - paving the way for new interdisciplinary research opportunities. By comparing our findings to a domain-specific benchmark, the Squire Index, we were able to show the potential and advantages of the application of process mining techniques in this domain. 

\textbf{Implications for Political Science Research:}
It has been discussed that while the Squire Index enables an easy and straightforward comparison, the insights it can deliver are limited due to its aggregated format \cite{bowen.2014}. Through our case study, we have shown that also process metrics can be considered when comparing state parliaments. Our findings show that the assumptions made by the Squire Index do not always fit the reality. In our analysis, we saw expected, unexpected, and the absence of significant relationships between process mining metrics and the Squire Index - suggesting a more complex relationship between legislative capacity and performance. 

Beyond what the Squire Index can express, our analysis revealed several noteworthy insights into the factors influencing legislative cycle times across the three state parliaments. Most notably, we found distinct patterns in what drives delays in each parliament: In Berlin, the key drivers appear to be particular delays between the first reading and committee consultations and, generally, between the first and second readings. However, also the process complexity in terms of event counts plays a role. In Brandenburg, high event counts and multiple committee sessions emerged as the strongest explanatory factors. Additionally, a great amount of all successful legislation ($82.3\%$ of passed bills) exceeded our cycle time threshold.
In Baden-Württemberg, the consistently lower recall values across most rules suggest that within this parliament, the factors driving slower processes (those exceeding its own average) may differ from what drives delays in the other parliaments.
These findings align with our earlier visual analysis and suggest fundamentally different approaches to legislative processes across the three parliaments. Brandenburg appears to have complex processes with multiple events and sessions, while Berlin's longer cycle times seem more attributable to delays between key process milestones. It is worth noting that many more insightful rules could be derived and tested similarly - for instance, directly checking for the influence of different political parties, or legislative topics. We left out these analyses due to space limitations.

\textbf{Implications for Cross-Organizational Process Mining:} We conclude that while we were indeed able to use process mining techniques in a real-world cross-organizational setting, there still remains a need for further research. First, our analysis assumed that all organizations willingly share their datasets - a condition that is often not realistic in industrial settings. This highlights the importance of developing cross-organizational process mining techniques that require less open data sharing than in our case. Second, we observed a need for more explicit tools for cross-organizational process mining - e.g., implementations of common techniques such as the dotted chart allowing for immediate side-by-side comparisons. 

Lastly, more research is needed to better integrate and understand the influences of properties of the business objects involved in processes - such as the complexity or controversy level of a law draft. It is reasonable to expect that the processes of very complicated and controversial law drafts differ substantially from those of simple law drafts with wide acceptance. For instance, the long waiting times we found for delayed processes in Berlin could be due to a high level of controversy or complexity. Benchmarking and performance measures should consider the diversity of the involved business objects for fair comparisons and to deliver trustworthy results. Therefore, process mining tools for benchmarking settings should enable the interactive exploration and identification of cohorts of processes in order to assess performance appropriately, for instance, in relation to the difficulty of the task.

\section{Conclusion}\label{sec:conclusion}

In this paper, we presented a case study on cross-organizational process mining. 
We demonstrated how current cross-organizational process mining techniques can be applied to compare the same process in different institutions.
In our case study, we analyzed the legislation processes of three German state parliaments and identified root causes for differences in their cycle time distributions. 
We showed that some differences can be explained through cross-organizational process mining that cannot be explained via traditional benchmarks such as the Squire Index.
Finally, to further harness the great potential of cross-organizational process mining, future research should examine the needs of practitioners and organizations through qualitative studies and develop improved benchmarking concepts based on the thereby obtained insights.

\section*{Acknowledgment}
The research of the authors was supported by the Einstein Foundation Berlin under grant EPP-2019-524, by the German Federal Ministry of Education and Research under grant 16DII133, and by Deutsche Forschungsgemeinschaft under grants 496119880 (VisualMine), 531115272 (ProImpact), and SFB 1404/2 (FONDA).
We thank Lena Ulbricht for her input on the political science part of our work. Further, we thank the FDP group of the 20th election cycle of the German Bundestag for discussing our method and results with us. 


\end{document}